\documentclass[preprint]{vgtc}               

\ifpdf
  \pdfoutput=1\relax                   
  \usepackage{graphicx}                
  \DeclareGraphicsExtensions{.pdf,.png,.jpg,.jpeg} 
\else
  \usepackage{graphicx}                
  \DeclareGraphicsExtensions{.eps}     
\fi%

\graphicspath{{figures/}{pictures/}{images/}{./}} 

\usepackage{microtype}                 
\PassOptionsToPackage{warn}{textcomp}  
\usepackage{textcomp}                  
\usepackage{mathptmx}                  
\usepackage{times}                     
\usepackage{cite}                      
\usepackage{tabu}                      
\usepackage{booktabs}                  

\usepackage{calc}
\usepackage{xcolor}
\usepackage{xspace}
\usepackage{multicol}
\usepackage{multirow}
\usepackage{tabularx}
\usepackage{booktabs}
\usepackage{colortbl}
\usepackage{amssymb}

\usepackage{siunitx}
\DeclareSIUnit[number-unit-product = {\thinspace}]{\year}{years}
\DeclareSIUnit[number-unit-product = {\thinspace}]{\inch}{inch}
\DeclareSIUnit[number-unit-product = {\thinspace}]{\second}{sec}

\definecolor{NoteBlue}{RGB}{0,0,191}

\definecolor{TodoRed}{RGB}{225,63,63}

\definecolor{ControversialGreen}{RGB}{0,127,0}

\definecolor{UnfinishedBlue}{RGB}{0,127,127}

\definecolor{KKOrange}{RGB}{255,103,2}

\definecolor{ChangelogPurple}{RGB}{255,20,128}
\newcommand\cl[1]{#1}

\newcolumntype{C}[1]{>{\centering\arraybackslash}p{#1}}
\definecolor{Gray}{gray}{0.85}
\newcolumntype{G}[1]{>{\columncolor{Gray}\centering\arraybackslash}p{#1}}

\definecolor{grayCell}{HTML}{d3d3d3}
\definecolor{greenP}{HTML}{bae4b3}
\definecolor{greenPP}{HTML}{74c476}
\definecolor{greenPPP}{HTML}{31a354}

\definecolor{barColor}{HTML}{ff7f00}
\definecolor{scatterColor}{HTML}{377eb8}
\definecolor{adaptedColor}{HTML}{4daf4a}
\definecolor{nonAdaptedColor}{HTML}{984ea3}

\usepackage{enumitem}
\setlist{nosep}

\newcommand{\fref}[1]{Fig.~\ref{#1}}

\newcommand{\afref}[1]{Fig.~A#1 in Appendix}
\newcommand{\atref}[1]{Tab.~A#1 in Appendix}
\newcommand{\asref}[1]{Sec.~#1 in Appendix}

\newcommand{\etal}[1]{{#1} et~al.\xspace}
\newcommand*{\parainline}[1]{{\fontsize{8pt}{8pt}\selectfont\sffamily\bfseries#1}\hspace{3pt}}

\newcommand{\vlLong}{visualization literacy\xspace}
\newcommand{\vl}{VL\xspace}
\newcommand{\dvlLong}{data visualization literacy\xspace}

\newcommand{\ueqLong}{User Experience Questionnaire Plus\xspace}
\newcommand{\ueq}{UEQ+\xspace}

\newcommand{\adap}{Adaptation\xspace}

\newcommand{\adapOn}{\textit{Adapted}\xspace}
\newcommand{\adapOff}{\textit{Non-Adapted}\xspace}

\newcommand{\visType}{Visualization Type\xspace}
\newcommand{\visTypes}{Visualization Types\xspace}
\newcommand{\visTypeBar}{\textit{Bar Chart}\xspace}
\newcommand{\visTypeBars}{\textit{Bar Charts}\xspace}
\newcommand{\visTypeScatter}{\textit{Scatter Plot}\xspace}

\newcommand{\taskCompTime}{\textit{task completion time}\xspace}

\newcommand{\taskAcc}{\textit{task accuracy}\xspace}

\newcommand{\ux}{\textit{user experience}\xspace}

\newcommand{\uxGroupA}{Dependability\xspace}
\newcommand{\uxGroupB}{Usefulness\xspace}
\newcommand{\uxGroupC}{Intuitive Use\xspace}

\newcommand{\taskCompTimeShort}{\textit{TCT}\xspace}
\newcommand{\taskAccShort}{\textit{TA}\xspace}

\newcommand{\participantMulti}[1]{{(#1)}}

\newcommand{\participantCount}[2]{{(#1 out of #2)}}
\newcommand{\participantPercentage}[1]{{(#1\%)}}

\newcommand{\msd}[2]{$M={#1}, SD={#2}$\xspace}

\newcommand{\anovaChi}[2]{$\chi^2(1)={#1}, p<{#2}$\xspace}

\newcommand{\lmmFixedFactor}[3]{$b={#1}, t={#2}, p<{#3}$}

\onlineid{1016}

\vgtccategory{Research}

\vgtcinsertpkg

\title{Who benefits from Visualization Adaptations? Towards a better Understanding of the Influence of Visualization Literacy}

\author{
    Marc Satkowski~\thanks{The first two authors contributed equally to this work.}~~\thanks{e-mail: [msatkowski, dachselt]@acm.org}\\
    \scriptsize Interactive Media Lab Dresden\\
    \scriptsize Technische Universit\"at Dresden\\ 
    \and
    Franziska Kessler~\footnotemark[1]{}~~\thanks{e-mail: [franziska.kessler, susanne.narciss]@tu-dresden.de}\\
    \scriptsize Psychology of Learning and Instruction\\
    \scriptsize Technische Universit\"at Dresden\\ 
    \and
    Susanne Narciss~\footnotemark[3]{}\\
    \scriptsize Psychology of Learning and Instruction\\
    \scriptsize Technische Universit\"at Dresden\\  
    \and
    Raimund Dachselt~\footnotemark[2]{}~~\thanks{Also with Clusters of Excellence Physics of Life and Centre for Tactile Internet with Human-in-the-Loop (CeTI), both at Technische Universität Dresden.} \\
    \scriptsize Interactive Media Lab Dresden\\ 
    \scriptsize Technische Universit\"at Dresden\\ 
}

\shortauthortitle{Satkowski \MakeLowercase{\textit{et al.}}: Towards a better Understanding of the Influence of Visualization Literacy}

\abstract{

The ability to read, understand, and comprehend visual information representations is subsumed under the term visualization literacy (VL).  
One possibility to improve the use of information visualizations is to introduce adaptations.  
However, it is yet unclear whether people with different VL benefit from adaptations to the same degree.
We conducted an online experiment (n = 42) to investigate whether the effect of an adaptation (here: De-Emphasis) of visualizations (bar charts, scatter plots) on performance (accuracy, time) and user experiences depends on users' VL level.
Using linear mixed models for the analyses, we found a positive impact of the De-Emphasis adaptation across all conditions, as well as an interaction effect of adaptation and VL on the task completion time for bar charts.
This work contributes to a better understanding of the intertwined relationship of VL and visual adaptations and motivates future research. 

}


\CCScatlist{
  \CCScatTwelve{Human-centered computing}{Visu\-al\-iza\-tion}{Visualization design and evaluation methods}{}
}

\begin{document}

\maketitle


\section{Introduction}

Visual representations of information pervade our everyday life and are already present at an early age in school and later during adulthood, on news websites, or on personal mobile devices.
The number of visualization types and the diversity of users is increasing continuously. 
Therefore, it is likely to see a wide variance in skill level across users as well as for different types of visualizations. 
Some users may be left behind and experience difficulties in understanding visualizations in general or have deficiencies in reading certain types of visual data representations.
The competence and the cognitive process related to the ability to read, understand, and comprehend visualizations have been summarized and conceptualized under the term \vlLong (\vl) \cite{Boy2014} or \dvlLong \cite{Borner2019}.
As the \vl level can differ between users and even visualization types for each user, one cannot take it for granted that specific visualization instances can be equally well understood by every person.

One possibility to support users is to adapt a given visualization to the specific characteristics of the current user, thus tailoring the presentation to their needs.
This in turn enhances the probability of conveying the information successfully.
There exist several aspects that can be adapted, such as
    changing visual channels \cite{Heer2010},
    using metaphors \cite{Ziemkiewicz2008}, as well as
    altering the layout or even completely changing the visualization type \cite{Ziemkiewicz2013}.
How adaptations of visualizations affect the performance and user experience is likely to be based on the \vl level of the individual user.
The aptitude-treatment interaction \cite{McLeod1978} describes the effect that the same instructional strategy (i.e., treatment, in this case the adaptation) can be more or less effective for individuals depending on their specific abilities.
Further, the expertise reversal effect \cite{Kalyuga2007} describes that instructional support exerts a positive effect on individuals with low level of prior knowledge, whereas the effect on experts can be detrimental. 
Hence, an adaptation can be beneficial for users if it matches the aptitude of the individual or detrimental in case of a low match.

In order to design the most conducive individualized visualization adaptations, it will be important to gain a better understanding whether the effect of specific visual techniques is dependent on individual \vl level.  
Therefore, we conducted an experiment with 42 participants to investigate the differential effect of visualization adaptations on performance and user experience in dependence of individual \vl levels. 
We used basic visualization types of bar charts and scatter plots and a simple highlighting and De-Emphasis \cite{Carenini2014} approach as an adaptation.
In the following, we will present 
    the design and results of our experiment\footnote{Study material and data are provided in the supplemental material.} on the effect of 2D visualization adaptation with regard to \vl level.

\section{Background \& Related Work}

This work touches the areas of adaptive information visualization as well as user characteristics, particularly \vlLong.

\subsection{Adaptive \& Responsive Information Visualizations}

Information visualizations can take on different forms, ranging from 
    basic visualizations (e.g., bar charts, scatter plots) \cite{Saket2018}, 
    to more complex ones (e.g., parallel coordinate plots, tree maps), 
    or specialized visualization types (e.g., Sankey charts) \cite{Poetzsch2020}.
In the presence of the growing population that interacts with an increasing number of visualization types, it becomes ever more challenging to create them in such a way that every person can equally read, interpret, and understand a given information visualization.
Dynamic adaptations are used to account for the individual differences and needs of users and thus aim to provide each individual with their ideal set of adaptations.
Responsive visualizations \cite{Choe2019, Lee2018, Lee2021} are a special type of adaptive data visualizations which are capable \textit{``to adapt themselves automatically to external contextual requirements''} \cite{Horak2021}.
Such a requirement could be the size of a device or the display resolution \cite{Choe2019, Lee2018}, data density, layout, and interaction-related aspects \cite{Hoffswell2020, Andrews2017, Lee2021}.
Adaptive visualization can also be based on explicitly provided or inferred user actions, characteristics, or other parameters \cite{Ahn2013}.
One way to facilitate the reading and interpretation of the depicted data is to, e.g., change visual marks or channels to alter the visual encoding \cite{Munzner2014} or a combination thereof.
Exemplary types of visual techniques that can be used for adaptations are 
    highlighting \cite{Toker2018}, 
    De-Emphasis \cite{Carenini2014}, 
    the introduction of additional visual overlay elements \cite{Ahn2013, Kong2012,Steichen2019}, or 
    simplifying \cite{Lalle2017, Yelizarov2014} 
    as well as hiding whole visualizations \cite{Lalle2017}.


\begin{figure}[!t]
    \centering
    \includegraphics[width=\columnwidth]{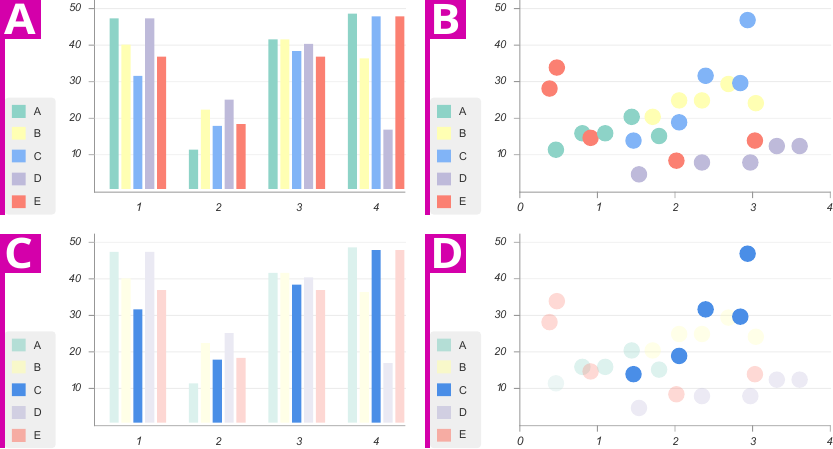}
    \caption{
        Schematic presentation of the used visualizations.
        \textbf{(A)} and \textbf{(C)} show a \visTypeBar, while \textbf{(B)} and \textbf{(D)} depict a \visTypeScatter.
        \textbf{(A)} and \textbf{(B)} show visualizations which are not adapted.
        On the other hand, \textbf{(C)} and \textbf{(D)}  show adapted visualizations using De-Emphasis.
    }
    \label{fig:visualizations}
    \vspace{-4mm}
\end{figure}

\subsection{User Characteristics \& Visualization Literacy}

There exists a growing body of related work that identifies internal and external user characteristics or properties \cite{Ahn2013} that can be used to trigger different adaptations. 
These are, e.g.,
    user context properties (e.g., education, aim) \cite{Golemati2006}, 
    cognitive load \cite{Yelizarov2014},
    personal traits (e.g., locus of control, cognitive style) \cite{Lalle2017, Carenini2014,Steichen2019}, 
    working and spatial memory \cite{Toker2018, Lalle2019}, or 
    areas of interest measured through gaze \cite{Lalle2017, Lalle2015, Conati2015,Steichen2013}.
The human ability to understand and comprehend visual stimuli can be described as visual literacy, the \textit{``ability to understand, interpret and evaluate visual messages"} \cite{Bristor1994} and consists of various dimensions for visual thinking, learning, and communication \cite{Aanstoos2004, Trumbo1999}.
Further visual competencies, like visual production, perception, interpretation, and reception \cite{Muller2008} can also be seen as parts of this literacy.
In the case of reading information visualizations, an extended skill set is required, which is defined as \vlLong (\vl) or \dvlLong (DVL) \cite{Borner2019} and can be described as \textit{``the ability to confidently use a given data visualization to translate questions specified in the data domain into visual queries in the visual domain, as well as interpreting visual patterns in the visual domain as properties in the data domain"} \cite{Boy2014}.
In order to base adaptation on \vl, it is vital to have a reliable and easily administered way of assessing this construct.
Existing approaches for \vl measurement include 
    the VLAT \cite{Lee2017}, 
    the DVL framework \cite{Borner2019}, and 
    the assessment of \vl based on Item Response Theory for line charts, bar charts, or scatter plots by \etal{Boy} \cite{Boy2014}.

Overall, the level of prior knowledge and competencies affect how well a task can be accomplished and how difficult and strenuous processing of the task will be perceived as by the users. 
Therefore, assistance and support in form of adaptations offered to the user are likely to have different effect on individuals with either high or low level of competence.
More precisely, additional support may not be beneficial in every case, but rather depends on the level of user expertise. 
\cl{It is observable that instructional support can have a detrimental effect on experts and a positive effect on users with low level of prior knowledge, which is further described by the expertise reversal effect \cite{Kalyuga2007, Kalyuga2003}.}
Adaptations are likely to be most beneficial if they match the users' individual competence level (i.e., \vl level) thus achieving an ideal aptitude-treatment interaction \cite{McLeod1978}.
Therefore, we conjecture that by taking the individual \vl level into account it may be possible to provide adaptations that are better suited to the user's requirements.

\section{Study Goals \& Hypotheses}
\label{sec:study:hypotheses}

Our goal is to shed light on the interaction of \vl and visualization techniques, which could be facilitated for adaptation in the future, on the users' performance and user experience.
Therefore, we conducted a study in which we manipulated the presentation state (de-emphasized and non de-emphasized; hereafter referred to as \adapOn and \adapOff) of two \visTypes (\visTypeBar and \visTypeScatter) in a randomized 2x2 factorial \cl{within-subject} design.
For this study, we generated three hypotheses: 
\begin{description}[style=multiline,leftmargin=0.7cm]
    \item[\textbf{H1}] 
        \textbf{(Main effect of \vl on task performance)}
        We expect participants with higher \vl to perform better than those with lower \vl scores.
    \item[\textbf{H2}]
        \textbf{(Interaction effect of \vl and \adap on task performance)}
        Higher performance is expected with \adapOn than with \adapOff visualizations, which difference should be more distinct for participants with lower \vl.
    \item[\textbf{H3}]
        \textbf{(Interaction effect of \vl and \adap on user experience)}
        We expect that user experience for \adapOn visualizations is rated more positively by participants with lower \vl compared to participants with higher \vl. 
\end{description}
All study materials, i.e., the complete questionnaire, the created visualizations, the tasks, as well as the collected study data and its analysis can be found in the supplementary material.

\subsection{Participants}

Our online experiment had a \cl{completion rate of \SI{38.4}{\percent}, which resulted in a total of 43 submitted and completed data sets.}
One data set was excluded as the subject did not comply with the instructions.
Promotion was done via mailing lists in our local university and over two survey websites\footnote{\url{https://surveyswap.io/} and \url{https://www.surveycircle.com}}.
All participants had a chance to win one of three 15€ Amazon vouchers.
As requested by the data security board of our local university we only recorded age groups.
Most of the 42 participants (19 female, 23 male) were in the age groups of 20 to 23 \participantCount{9}{42}, 24 to 27 \participantCount{18}{42}, and 28 to 31 \participantCount{9}{42}, the remaining were older than 31 \participantCount{6}{42}.
All participants reported an academic background.
Further three indicated to have a red-green weakness.

\subsection{Task Design}
\label{sec:study:tasks}

\parainline{Question Design} 
We created questions based on the low-level analysis task taxonomy of \etal{Amar} \cite{Amar2005}.
Concretely, we decided to use the low-level tasks of \textit{Filter}, \textit{Determine Range}, and \textit{Compute Derived Value} and combined two of those for every question  (see \atref{1}).
We created five task groups, which were repeated in each condition. 
Each question in a task group was based on the same structure wherein only specific values of the respective data attributes were altered (e.g., country names, year).
We created a total of 20 questions (5 task groups x 2 visualization types x 2 adaptation styles).
\newline 
\parainline{Design of Visualization Condition} 
We constrained our study to two \visTypes (\visTypeBar and \visTypeScatter).
We used the De-Emphasis approach presented by \etal{Carenini} \cite{Carenini2014} in the \adapOn condition \cl{(see \afref{1})}, as it creates a simple pop-out effect \cite{Munzner2014} thus highlighting important data points. 
20 visualizations were created (see \fref{fig:visualizations}) \cl{using features presented in} Tableau\footnote{\url{https://www.tableau.com/}}, ten for each \visType (five \adapOn and five \adapOff).
Each of the \adapOn visualizations were handcrafted based on their corresponding question.
All 20 visualizations were based on the same data set generated from gapminder\footnote{\url{www.gapminder.org} and \url{https://public.tableau.com/en-us/gallery/how-has-world-changed-1962}}.
In all visualizations, groups of e.g., years, were visually separated by the use of different colours. 
Grid lines were included in the visualizations in order to make it easier to read values from the axes (see \fref{fig:visualizations}).

\subsection{Data Collection \& Measurement}

\parainline{Visualization Literacy Assessment} 
We used the \vl assessment of \etal{Boy} \cite{Boy2014} to record the individual \vl level of each participant.
For the assessment, participants were redirected to the online version of the test\footnote{The online version is no longer accessible but code is still available here: \url{https://github.com/INRIA/Visualization-Literacy-101}}.
The test measured \vl scores separately for \visTypeBar and \visTypeScatter.
\newline 
\parainline{Task Performance} 
The task performance was operationalized as the \taskCompTime (\taskCompTimeShort) and \taskAcc (\taskAccShort).
Each data point greater than $M + 2 * SD$ was defined as an outlier and was subsequently replaced by the exact value of this formula, as proposed by \etal{Field} \cite{Field2012}.
A total of \SI{4.4}{\percent} values were classified as outliers and replaced.
We used multiple-choice tests with up to seven options for the tasks, whereof only one answer option was correct.
Scoring for the \taskAccShort was mapped to 1 if the answer was correct and to 0 if the answer was incorrect.
\newline 
\parainline{User Experience} 
To measure \ux, we used three scales from the \ueqLong (\ueq)\footnote{English version: \url{http://ueqplus.ueq-research.org/}} \cite{Schrepp2019,Schrepp2017,Laugwitz2008}: \uxGroupA, \uxGroupB, and \uxGroupC.
Each scale contains four questions on a seven-point likert scale.
Additionally, we asked the participants to state whether they preferred the \adapOn or \adapOff version of both \visTypes. 

\subsection{Setup \& Procedure}

The study was conducted as an online experiment implemented in LimeSurvey\footnote{\url{https://www.limesurvey.org/}}.
It consisted of the following parts:
    (1) A demographic questionnaire; 
    (2) \vl assessment \cite{Boy2014} for \visTypeBar, followed by \visTypeScatter;
    (3) the \adapOff tasks;
    (4) the \adapOn tasks;
    lastly, (5) a post-study questionnaire focused on user preferences and procedures.
To reduce a potential carry-over and anchoring effects, we decided to present tasks on \adapOff visualization (3) before the \adapOn visualizations (4), while the order of the five tasks within each adaptation block (3 and 4) was randomized.
Participants were asked to report their \vl assessment \cite{Boy2014} score obtained on their return to the online experiment site.
After each \adap condition in (3) and (4) the participants were asked to answer the \ux questionnaire \cite{Schrepp2019}.
The total duration of the experiment averaged to approximately \SI{50}{\min} (\msd{\SI{50.26}{\min}}{\SI{14.57}{\min}}) while around \SI{12}{min} (\msd{\SI{12.24}{\min}}{\SI{4.47}{\min}}) were needed for the \visTypeBar and for the \visTypeScatter \vl assessment each.

\section{Data Analysis \& Results}

We will describe the main data analyses and findings.
Additional analyses focusing on the collected \vl scores \cl{(see Sec. A and \afref{2})} and tables (see Tab.~A2 to A5 in Appendix) are provided with the supplemental material.


\begin{figure*}[!t]
    \centering
    \includegraphics[width=\textwidth]{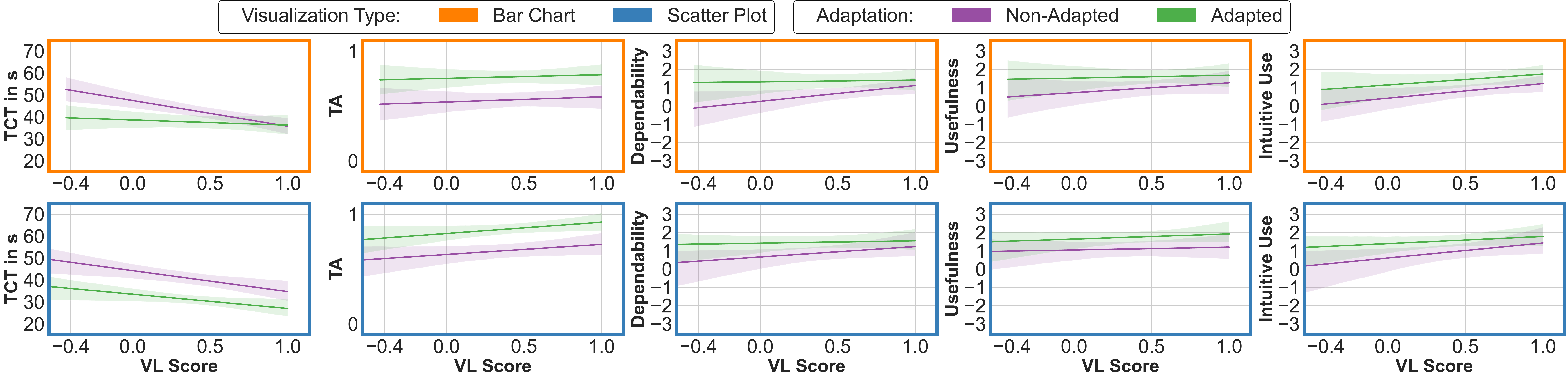}
    \caption{
        The linear regression of all dependent variables (\taskCompTime{} as \taskCompTimeShort{} and \taskAcc{} as \taskAccShort{}) over the \vl.
        \textcolor{barColor}{\textbf{\visTypeBar{}}} and \textcolor{scatterColor}{\textbf{\visTypeScatter{}}} are presented in different plots, while \textcolor{adaptedColor}{\textbf{\adapOn{}}} and \textcolor{nonAdaptedColor}{\textbf{\adapOff{}}} are presented as different lines.
        The shadow behind the lines shows the confidence interval of 95\%. 
        }
    \label{fig:results}
    \vspace{-4mm}
\end{figure*}

\subsection{Data Analysis Methods}

We used JASP\footnote{\url{https://jasp-stats.org/}} for \cl{the} data analyses. 
We performed the following statistical tests on both \visTypes independently.
We applied multilevel modeling \cite{Field2012,Winter2013,Singmann2019,Greenland2000} in order to account for the repeated measures and \vl score interaction.
We constructed a linear mixed model for \taskCompTime and a generalized linear mixed model (GLMM) \cite{Baayen2008} for the dichotomous variable \taskAccShort.
In order to test main and interaction effects of the fixed effect factors (\vl and \adap), we used the likelihood test ratio method to compare the crossed random effect models.
In order to account for the violation of independence of the repeated measurements as well as the measurements of the task groups that are expected to be more similar within one task group, we included the factors participants, task groups (see \atref{1}), and their interaction as random effects into the model (for \taskCompTimeShort and \taskAccShort).
In the model for \ux values, the interaction term was dropped as this parsimonious model provided a better fit. 


\subsection{Results}

\parainline{Task Completion Time} 
For the \taskCompTime (\taskCompTimeShort), we used a linear mixed model (see \fref{fig:results}) without random slopes as it showed the best model fit.
For \visTypeBar, we found a relationship between the \vl (\anovaChi{4.009}{.05}), \adap (\anovaChi{20.432}{.001}), and their interaction (\anovaChi{8.882}{.01}) on the \taskCompTimeShort across the participants and task groups (see \atref{2}).
This shows that the \taskCompTimeShort can be significantly predicted by the \vl (\lmmFixedFactor{-7.046}{-2.052}{.05}), by the \adap (\lmmFixedFactor{-4.43}{-4.583}{.001}), and the interaction of both (\lmmFixedFactor{4.689}{2.998}{.01}) (see \atref{4}).
In contrast, for \visTypeScatter, we found only a relationship between \vl (\anovaChi{6.365}{.05}) and \adap (\anovaChi{40.867}{.001}) on the \taskCompTimeShort, but no interaction effect (see \atref{3}).
This in turn shows that the \taskCompTimeShort can be significantly predicted by the \vl (\lmmFixedFactor{-8.032}{-2.623}{.5}) and the \adap (\lmmFixedFactor{-5.368}{-6.571}{.001}) (see \atref{5}).
\newline 
\parainline{Task Accuracy} 
For the \taskAcc (\taskAccShort), we used a generalized linear mixed model (see \fref{fig:results}) of the binomial family (logit link\footnote{In the logit model the log odds of the outcome is modeled as a linear combination of the predictor variables.}) and a random slope for \adap. 
For \visTypeBar, we did not find any relationship between the fixed effects and the \taskAccShort (see \atref{2}).
For \visTypeScatter, we found a relationship between the \adap (\anovaChi{12.774}{.001}) and the \taskAccShort (see \atref{3}).
This in turn shows that the \taskAccShort can be significantly predicted by the \adap (\lmmFixedFactor{0.641}{4.126}{.001}) (see \atref{5}).
\newline 
\parainline{User Experience Ratings} 
For all three \ux scales, we used a linear mixed model (see \fref{fig:results}) without random slopes as they showed the best model fit.
For \visTypeBar, we found a relationship between the \adap and the \uxGroupA (\anovaChi{12.65}{.001}), \uxGroupB (\anovaChi{5.429}{.05}), and \uxGroupC (\anovaChi{6.16}{.05}) but no effects of \vl and no interaction effect (see \atref{2}).
This in turn shows that the \adap can significantly predict the \uxGroupA rating (\lmmFixedFactor{0.535}{3.842}{.001}), \uxGroupB rating (\lmmFixedFactor{0.399}{2.407}{.05}), and \uxGroupC rating (\lmmFixedFactor{.362}{2.576}{.05}) (see \atref{4}).
The same holds true for \visTypeScatter, where we found a relationship between the \adap and the \uxGroupA (\anovaChi{16.254}{.001}), \uxGroupB (\anovaChi{7.579}{.01}), and \uxGroupC (\anovaChi{14.832}{.001}) (see \atref{3}) showing that the \adap can significantly predict the \uxGroupA rating (\lmmFixedFactor{0.378}{4.455}{.001}), \uxGroupB rating (\lmmFixedFactor{0.298}{2.882}{0.01}), and \uxGroupC rating (\lmmFixedFactor{0.393}{4.218}{.001}) (see \atref{5}). 
We found no effects for \vl and no interaction effects. 
\newline 
Our data showed that the participants (P) slightly preferred the \adapOn visualizations over the \adapOff ones, for both the \visTypeBar{} (55\% of all participants) and the \visTypeScatter{} \participantPercentage{62}. 
This was further supported by participants' comments, indicating that the De-Emphasis approach helped them to focus on the given task \participantPercentage{48}. 
Some participants reported a beneficial effect of the color \participantPercentage{38}, e.g., \textit{``different colors [helped] to better differentiate [data points]''} \participantMulti{P1}.
Participants also acknowledged the beneficial effects of the adaptations by reducing the overwhelming amount of information \participantPercentage{24}.
Specifically, one participant labeled \textit{``graphs where too much data was presented simultaneously''} \participantMulti{P41} as frustrating.
Some participants also highlighted that the tasks were challenging \participantPercentage{17}.
One participant experienced \textit{``understanding the question and searching for the applicable bars/dots''} \participantMulti{P34} as frustrating.

\subsection{Hypotheses Results}
\label{sec:study:hypothesesResults}

\parainline{(H1)} 
We found a significant effect of \vl on the \taskCompTime (\taskCompTimeShort) where higher levels of \vl were associated with lower \taskCompTimeShort for \visTypeBar and \visTypeScatter (see \fref{fig:results}).
However, for the performance indicator \taskAcc (\taskAccShort), we did not find a significant effect of \vl.
Therefore, \textbf{H1} was only partly confirmed as only time was affected by \vl level but not accuracy. 
\newline 
\parainline{(H2)} 
We found that \adap had a positive effect on both, the task performance and the \ux (see \fref{fig:results}).
In general, participants working with \adapOn visualizations had a lower \taskCompTimeShort.
Further, we only found a significant positive effect of adaptations on the \taskAccShort for \visTypeScatter.
We saw an interaction effect on the \taskCompTimeShort for \visTypeBar (see \fref{fig:results}), indicating that participants with lower \vl might benefit more from the presented adaptations than participants with higher \vl.
However, we did not find any interaction effects for \taskAccShort.
Hence, \textbf{H2} was only partly confirmed.
\newline 
\parainline{(H3)} 
Participants working with \adapOn visualizations reported higher ratings in the three scales of \ux for both \visTypes (see \fref{fig:results}). 
However, we did not find any interaction effect of \vl and \adap for any \ux ratings, which shows that a simple visualization technique (i.e., De-Emphasis), appears to be beneficial for all participants with regard to \ux.
Further, we did not find any detrimental effect of \adap for higher levels of \vl.  
Therefore, we reject \textbf{H3}.

\section{Discussion}

We found support for the notion that the effect of \adap varies for different levels of \vl, in form of an interaction effect of \adap and \vl on \taskCompTimeShort for \visTypeBars. 
However, we did not find any other interactions.
We conjecture that the visualization technique De-Emphasis, which reduces the amount of information presented to the users, has a positive effect for users across different \vl, explaining the results with regard to hypotheses \textbf{H2} and \textbf{H3}.
\cl{On the other hand, it} is conceivable that adaptation strategies that present additional visual elements \cl{with the aim to support the understanding of the visualization} may be of hindrance for more experienced users.
\cl{Examples of such adaptation strategies are provision of tooltips, additional visual elements such as arrows, or textural hints.}
As external information and internal knowledge from the long-term memory needs to be integrated, the additional information that may be redundant for experts employs additional strain on the working memory of experts, thus increasing cognitive load instead of reducing it and consequently diminishing experts’ performance \cite{Kalyuga2007}.
Meanwhile the additional information may facilitate the understanding of more complex visualizations for inexperienced users, i.e., with lower \vl.

We could see differences for both \visTypes in the participants' familiarity rating, task performance, and \ux.
It is conceivable that the De-Emphasis approach is more effective on bar charts, since bars make up a larger portion of the diagram than the points in a scatter plot.
These results show that the quality and the layout of visual marks influence the effectiveness of a given adaptation.
Other visualization types use different features (e.g., lines) and can additionally be less common, like parallel coordinate plots or tree maps, which may result in different effects of adaptations.

The overall \vl level within our sample seemed to be above average (see \asref{A}).
One reason for this could be the academic background of the participants in our sample \cite{Maltese2015}.
Since visualizations play a vital role in academic teaching and thinking, it is likely that academics are more accustomed to and practiced at dealing with information visualizations resulting in high levels of \vl. 
Therefore, we can only draw conclusions about the interplay of \vl and adaptation for a limited range of \vl scores, which in turn affected hypotheses \textbf{H2} and \textbf{H3}.
Additionally, we think that the quality of the \vl assessment should be improved by 
    exploring benefits of aggregated (e.g., VLAT \cite{Lee2017}) or separate assessments for different visualization types (e.g., \vl assessment \cite{Boy2014}), as well as
    generating reference group baseline \vl scores.
\cl{Lastly, as we found an effect of \vl on \taskCompTimeShort, which can be partly explained by the type of \vl assessment (e.g., Item Response Theory for \etal{Boy} \cite{Boy2014}), we believe that new types of test could even map other user performance and experience properties besides the \taskCompTimeShort.
}

\section{Conclusion}

In this work, we investigated the effect of visualization strategies, i.e., De-Emphasis, on bar charts and scatter plots with regard to the user characteristic \vlLong (\vl).
Our findings suggest that taking individual \vl levels into account may be a promising way to create adaptations tailored to the individual needs.
Further research is required to substantiate the effect for other types of \cl{visualizations and adaptations}. 
We hope that our work can be used as a stepping stone in future research on adaptive visualizations based on \vl.


\section*{Acknowledgments}{
    We thanks Vincent Schmidt for the study support.
    This work was funded by the Deutsche Forschungsgemeinschaft (DFG)
        by DFG grant 319919706/RTG~2323, 
        by DFG grant 389792660 as part of TRR~248 -- CPEC (see \url{https://perspicuous-computing.science}),
        and as part of Germany's Excellence Strategy – EXC-2050/1 – 390696704 – Cluster of Excellence \textit{``Centre for Tactile Internet with Human-in-the-Loop'' (CeTI)} of TU Dresden.

}

\bibliographystyle{abbrv-doi}


\bibliography{main-v2-shortest}
\end{document}